\documentclass[11pt,a4paper]{article}
\pdfoutput=1

\usepackage[english]{babel}
\usepackage{epsfig}

\usepackage[utf8]{inputenc}
\usepackage{floatrow}
\usepackage{amssymb,amsmath,amsfonts,amsthm,graphicx,psfrag}
\usepackage{indentfirst}
\usepackage{hyperref}

\usepackage[title,titletoc]{appendix}
\usepackage{graphicx}
\usepackage{amsfonts}
\usepackage{bm}
\usepackage{verbatim}
\usepackage{epsfig}

\graphicspath{{./pictures/}}

\usepackage{mathrsfs}
\usepackage{amsthm}
\usepackage{amssymb}
\usepackage[T1]{fontenc}

\usepackage{authblk}
\usepackage{cite}
\usepackage{slashed}
\usepackage{amsmath}
\usepackage[left=2cm,right=2cm,
    top=2cm,bottom=2cm,bindingoffset=0cm]{geometry}
\newcommand{\be}{\begin{eqnarray}}
\newcommand{\ee}{\end{eqnarray}}

 \def\la{\mathrel{\mathpalette\fun <}}
\def\ga{\mathrel{\mathpalette\fun >}}
\def\fun#1#2{\lower3.6pt\vbox{\baselineskip0pt\lineskip.9pt
\ialign{$\mathsurround=0pt#1\hfil ##\hfil$\crcr#2\crcr\sim\crcr}}}

\newcommand{\veS}{\mbox{\boldmath${\rm S}$}}

\newcommand{\veB}{\mbox{\boldmath${\rm B}$}}

\newcommand{\lan}{\langle}
\newcommand{\ran}{\rangle}

\usepackage{authblk}

\title{Thermodynamics of a quark-gluon plasma at  finite baryon density.}
\author[+,*]{Z.V Khaidukov}
\author[*]{Yu.A.Simonov}
\affil[+]{Moscow Institute of Physics and Technology,
Institutskiy pereulok 9, 141700 Dolgoprudny, Moscow Region, Russia}
\affil[*]{Institute for Theoretical and Experimental Physics,NRC "Kurchatov Institute", B. Cheremushkinskaya 25, Moscow, 117259, Russia}

\begin{document}
  \maketitle
\begin{abstract}

Properties of the quark-gluon plasma(QGP) in   the presence of  the baryon chemical potential  $\mu_B$ are studied
using the Field Correlator Method(FCM). The  non-perturbative FCM dynamics includes the Polyakov line,  computed via colorelectric string tension $\sigma^E(T)$ and  the  quark and gluon Debye masses, defined by the colormagnetic string tension $\sigma^H(T)$.     The resulting   QGP thermodynamics at $\mu_B \le 400$ MeV is in a  good agreement with the available  lattice data,both pressure and the sound velocity do not show any sign of a 
critical behavior in this region.

\end{abstract}

\section{Introduction}

The main result of heavy ion experiments, performed over the last 15 years at RHIC and then at RHIC and LHC,  is the discovery of a new form of matter \cite{QGP1,QGP2,QGP3,QGP4,QGP5} with its properties markedly different from the pre-RHIC era predictions, see \cite{HIC1,HIC2,HIC3,HIC4,HIC5,HIC6,HIC7,HIC8,HIC9,HIC10} and references therein.  Instead of the commonly assumed picture of a  weakly interacting Quark-Gluon Plasma(QGP),  one possibly   has     a strongly coupled liquid  subject  to the  law of   the relativistic  hydrodynamics \cite{Hydro1,Hydro2,Rev}.The properties  of the produced matter  drastically  change  during   several stages  of evolution: from the stage of  formation,   hydrodynamization and thermalization towards the hadron gas   production. The wealth of the QCD matter phases is reflected in the QCD phase diagram drawn in the ($\mu,T$) plane. However, the correspondence between the specific($\mu,T$) domains of the phase diagram and the space-time dynamics of the fireball should be considered with caution.
The  reason is that the phase diagram describes the limit of  an infinite system in thermodynamic equilibrium.\par

From  the theoretical viewpoint  the matter created in heavy ion collisions should be described by the fundamental laws of QCD.  
For   these reasons  the  dynamics and thermodynamics of QCD at finite temperatures  is now in the focus of numerous investigations. At this moment  one of the  main sources of information is  the lattice calculations.  The  presence of strong interaction in QGP at zero baryon density was demonstrated in  numerous studies  \cite{qlatt1,qlatt2,latt1,latt3,latt4,lattglu4}. They show that the ratio of the QGP pressure to the non-interacting  case is less than 0.8 and  remains almost constant with  increasing temperature. \par 
 
Another striking discovery in this domain was the analysis of the temperature transition, made in the $2+1$ QCD lattice  computations, which has shown a smooth crossover in the temperature region $T=140\div 180$ MeV \cite{25*}\footnote{This QCD crossover  is a new phenomenon, possibly having some analogues in the material sciences  and in the ionization  and  dissociation  processes. The question of the existence of a critical point at  finite baryon  chemical potential is still of intense interest \cite{Crit}.}

Despite  a dramatic progress  the question about the  structure  of the QCD  phase diagram at  nonzero baryon density remains open.
This  happens mostly  because lattice methods  are  strongly restricted   to a  domain of small chemical potentials\footnote{We want to point out that   another very important  source of information of QCD phase diagram is connected with neutron stars physics \cite{Blinnikov}.     LIGO and Virgo's discovery of gravitational waves from neutron star \cite{LV}
opened a new  era of quark matter studies.Possible discovery of quark stars will give much more opportunities for the QCD phase diagram studying \cite{Itoh,Witten}.  } ($N_c$=3)  due to the ``sign problem''.  To circumvent this difficulty  in the case of  $N_{c}=3$    one   can use   the  Taylor  expansion  around zero chemical potential  \cite{thLatt1,thLatt2}, or  use   imaginary  chemical potential\cite{thLatt3}. Another possibility is to decrease   the number of colours to $N_c=2$, where the sign problem is absent \cite{Kot1,Kot2,Kot3,Kot4}.
\par

From all these facts      the need for analytic   methods  that can help with  investigation of  QGP thermodynamics and QCD phase diagram becomes obvious.
In this paper we will focus on   the Field Correlator Method (FCM), which  is applicable  in  QCD   at any chemical potential and any temperature \cite{VCM2,VCM3,VCM4,VCM5,VCM6,VCM7}. In this method    the   non-perturbative  dynamics in  confinement and deconfinement regions is based on vacuum properties, described by gluonic field correlators \cite{VCM2,VCM3,VCM4,VCM5,VCM6,VCM7,3,19,20,21,22,23,24,25,26} and  the key role     is played by   correlators  of colorelectric fields $D^E$ and colormagnetic  fields $D^H$, which provide colorelectric confinement (CEC)  with the string tension $\sigma^E (T)$ and colormagnetic  confinement (CMC) with the string  tension $\sigma^H (T)$.  
The latter  being   calculated   from field  correlators and  on the  lattice, grows with $T$ , $\sigma^H (T) \sim g^4 (T) T^2$ and insures the strong interaction at large $T$ mentioned above. 
It is   interesting to note, that in the  FCM the crossover phenomenon     is  connected with  the gradual vanishing of the   vacuum confining correlator $D^E(z)$ (and the resulting string tension $\sigma^E (T)$) with  the growing temperature. The same phenomenon of the ``melting  confinement'' can be observed in the SU(3) gluodynamics \cite{23},  where   the  string tension $\sigma(T)$, measured on the lattice \cite{45,46,47,48}  is also decreasing with $T$,   but in the case of SU(3) it cannot smoothly match the fast growing gluon pressure (in contrast to the slowly  growing glueball pressure due to  large glueball masses $\ga 2$ GeV). As a result, one has a weak  first order transition in SU(3)  \cite{23},
while in the $n_f = 2+1$ QCD with low mass  mesons  the  smooth matching of pressure is achievable in the course of transition.\par

As a proof of this picture one can use  the quark condensate vanishing with   $T$ \cite{qlatt1,latt3,Baz} in the same way, as $\sigma (T)$ (see e.g. Fig.4 in \cite{45},Fig.6 in \cite{47} with Fig.4 in \cite{qlatt1}),  which  is  connected with   confinement via  $\lan \bar qq \ran \sim \sigma^{3/2}  $ \cite {49,50,51}, and almost the same reasoning can extend this connection to nonzero $T$.\par

One can see many  important questions  in QCD that could be investigated by the FCM, and  we will focus on one of them: the  main task of this paper is  to give  a self-consistent description of  QGP at nonzero $\mu_B$.  

 We shall use below the thermodynamic
formalism exploited before for the gluon plasma in\cite{22,23,24 }  and extended
to the QGP case in Refs. \cite{25,26}.

 \par The paper is organized as follows:  In section 2 we introduce the FCM in  the case of finite temperature and  chemical potential.  In     section 3 we calculate the Polyakov line(\textit{L(T)})     in the case of (2+1)$n_f$ QCD.  In section 4  we extend the FCM formalism to nonzero $\mu_B$.  In section 5 we compare   our results with   lattice data  at zero and finite baryon chemical potential. Section 6   is devoted to  conclusions and the outlook. \par

\section{The Field Correlator Method at Finite temperature}

  The field correlator method(FCM)  is very powerful tool for describing physics of QCD(see \cite{**}  for a recent review) which allows to formulate "confinement"  or in other words    to obtain the area law of the Wilson loop  in terms of the vacuum background fields, with the field correlators  $D^{E},D^{H}$ ensuring colorelectric and colormagnetic confinement  with the string tensions  $\sigma^{E}$ and $\sigma^{H}$.As a result all hadron masses are defined in this method only by fermion masses and string tension $\sigma^{E}$. \par

All gluon fields $A_{\mu}$ in QCD in the  framework of the  background perturbation theory \cite{60*} can be divided into vacuum background  part $B_{\mu}$ and perturbative part $a_{\mu}$,
$A_{\mu}=B_{\mu}+a_{\mu}$, with $B_{\mu}$ contributing to $\sigma^{E}, \sigma^{H}$  while  $a_{\mu}$ is treated in the  background perturbation theory  with the perturbative  coupling constant $\alpha_{s}(Q)$, defined by the scale parameter $\Lambda_{QCD}$.\par

For the hadron spectrum in QCD  and for the QCD thermodynamics the basic role is played by the background fields $B_{\mu}$, while $a_{\mu}$ yield  perturbative corrections. On the other hand  in high momentum processes with $Q^{2}\gg M^{2}_{B}=2\pi\sigma=O(1$ GeV) the basic role is played by the perturbative  fields $a_{\mu}$. The boundary $M^{2}_{B}$   found in \cite{60**}   separates  both types of dynamics and $\sigma$ itself  defines  the scale $\Lambda_{QCD}$ \cite{**}. In this sense  the fields $B_{\mu}$ and $a_{\mu}$ can be associated with  to the regions $Q^{2}\le M^{2}_{B}$ and $Q^{2}> M^{2}_{B}$ respectively. \par

In thermodynamics at temperatures  $T\le M_{B}$ the basic dynamics is given by the background fields $B_{\mu}$ which define both colormagnetic confinement(yielding CMC Debye screening) and Polyakov line interactions. In what follows we shall concentrate on these contributions, taking into account  gluon exchange corrections. The  fundamental role in FCM is played by the  quadratic gluonic field correlator. It consists of two terms  $D $ and $D_1$

\be
D_{\mu\nu\lambda\rho}=g^{2}tr_{a}<F_{\mu\nu}(x)\Phi(x,y)F_{\lambda\rho}(y)\Phi(y,x)>=(\delta_{\mu\lambda}\delta_{\nu\rho}-\delta_{\mu\rho}\delta_{\nu\lambda})D(x-y)+\frac{1}{2}\left[ \frac{\partial}{\partial_{\mu}}(x_{\lambda}\delta_{\nu\rho}-x_{\rho}\delta_{\nu\lambda}) \right.\\ 
+\left.(\mu\lambda\leftrightarrow \nu\rho)\right]D_1 (x-y). \label{correl} \nonumber
\ee
 Here the parallel transporter $\Phi(x,y) =P\exp (ig \int^x_y du_\nu A_\nu (u)),$ and 
  the fields  $F_{i4},F_{4i}$ refer to $D^{E},D^{E}_{1}$ and $F_{ik}$ to the  $D^{H},D^{H}_{1}$ correlators.
One can obtain string tension via $D^{E},D^{H}$:
\be
\sigma^{E,H}=\frac{1}{2}\int{D^{E,H}}d^{2}z  
\label{2.2}\ee

At zero temperature  $\sigma^{E}=\sigma^{H}$. 
Let's discuss in more detail the basic principles of FCM at finite temperatures.  
We must take into account that at finite temperatures the  confinement-deconfinement transition occurs. In our formalism that means that electric string (or colorelectric correlator $D^E$) has to  vanish. But there is no  restrictions  on  the value of   colormagnetic  correlator (or alternatively  on  the existence of  colormagnetic string  tension $\sigma_H$). As shown by analytic \cite{42***} and lattice studies  $\sigma_H$ grows quadratically with  temperature. As a result for $T>T_{c}$   there is no  confining string  between colour charges, but there  is  still non-perturbative interaction between them i.e.  colorelectric (CE) interaction, contained in the Polyakov line L(T), and the colormagnetic  confinement (CM)   in a   spatial projection of the Wilson loop. 
  Analysis of  physics of QGP  in terms of FCM   made in \cite{3,19,20,21,TVCM1,TVCM2}, also confirmed   the important role of Polyakov loops   for description of thermodynamic of GP and QGP. In  \cite {22,23,24,25,26} also the CMC interaction was taken into account, providing a selfconsistent dynamical picture in a good agreement with lattice data.
 As for CMC   it is  the   main interaction in QGP, operating   above transition temperature, as was observed in lattice data \cite{42*}, where the CMC correlators $\lan tr F_{ik} (x) \Phi (x,y) F_{ik} (y)\ran$ have been measured, see also \cite{63*} where $\sigma^H$ was  studied on the lattice, and  \cite{42***}  where $\sigma^H$ was estimated in FCM.

It was found in \cite{42**} that CMC does not support white bound states in $q\bar q$ and $gg$  systems at zero temperature, however it can create the screening mass $M(T)$ of isolated quarks and gluons \cite{22,23,24,25,42***}, which grows with temperature so that the ratio $\frac{M(T)}{T}$ is constant up to the logarithmic terms.  \par

As it was shown in \cite {3,19,20,21, 22,23,24,25,26} the    most convenient for description of  QCD thermodynamics  is the T-dependent path integral (worldline) formalism, where   pressure  can be written in the form\cite{3,21,23}  (see  Appendix 1 for details of derivation)
 \be
 P_{gl}=2(N^{2}_{c}-1)\int_{0}^{\infty}\frac{ds}{s}\sum_{n=1,2..}G^{n}(s).\label{2.3}
\ee
Here s is the proper time, and for $G^{n}(s)$ one can obtain:
\be
G^{n}(s)=\int{(Dz)^{\omega}_{on}exp(-K)\hat{tr}_{a}<W^{a}_{\Sigma}(C_{n})>}, \label{2.4}
\ee
where $K=\frac{1}{4}\int^{s}_{0} d\tau\left(\frac{dz^{\mu}}{d\tau}\right)^{2}$, and $W^{a}_{\Sigma}(C_{n})$ is the adjoint  Wilson loop
defined for the gluon path $C_{n}$, which has both temporal (i4) and spacial projections (ij), and $\hat{tr}_{a}$  is the normalized
adjoint trace.   When $T>T_{c}$  the correlation function between CE and CM fields is rather week \cite{3}:
\be
\lan E_{i}(x)B_{k}(y)\Phi(x,y)\ran\approx 0 \label{2.5}
\ee  

At this point when averaging $W^a_\Sigma (C_n) $ in (\ref{2.4}) one should take into account that the paths of gluons at $n_0 \neq 0$ are not closed, and there is a free piece of $n$ temporal steps, which should be  connected by the gluon path to form a closed  contour of Wilson loop, with the area law in the vacuum  confining field. Therefore one can  add before vacuum  averaging a piece along  time axis, which closes the gluon trajectory from $n_0$ to $n=0$ and  back from $n_q=0$ to $n=n_0$ (which is an identical operation yielding a factor 1). In this way one obtains a product of a closed contour and a Polyakov line from $0$ to $n_0$, and the vacuum averaging yields the 
  expression for the factorized  Wilson loops   \cite{23}:
\be  <W^{a}_{\Sigma}(C_{n})>=L^{(n)}_{adj}(T)<W_{3}>\label{2.6} \ee 
 with  $L^{(n)}_{adj} \approx L^{n}_{adj}$ for $T \le 1$ GeV.
One can integrate out  the $z_{4}$ part of the path integral $(Dz)^{\omega}_{on}=(Dz_{4})^{\omega}_{on}D^{3}z$, with the result 
\be G^{(n)}(s)=G^{(n)}_{4}(s)G_{3}(s)  , ~~ G^{n}_{4}(s)=\int (Dz_{4})^{\omega}_{on}e^{-K}L^{(n)}_{adj}=\frac{1}{2\sqrt{4\pi s}}e^{-\frac{n^{2}}{4T^{2}s}}L^{(n)}_{adj} \label{2.7}\ee
This  factorization holds also  for quarks and will be used below (with  the change of   the adjoint representation to    the fundamental one).\par

The resulting gluon  contribution is 
\be
P_{gl}=\frac{2(N^{2}_{c}-1)}{\sqrt{4\pi}}\int^{\infty}_{0}\frac{ds}{s^{3/2}}G_{3}(s)\sum_{n= 1, 2,...}e^{-\frac{n^{2}}{4T^{2}s}}L^{n}_{adj}, 
G_{3}(s)=\int (D^{3}z)_{xx}e^{-K_{3d}}<\hat{tr}_{a}W^{a}_{3}> \label{eqg}
\ee 

The direct appearance of $L^{(n)}_{adj}(T)$ in the  thermodynamic potential is an important feature of the present non-perturbative formalism based on FCM. It was  derived before in \cite{3}, when the CMC was  not taken into account, and the origin of $L^{(n)}$ was associated only with the correlator $D^E_1$ \cite{TVCM1}. As will be shown below the mechanism of the Polyakov loop is  much more  complicated, and we shall compute $L(T)$ in a  different way.

At this point we are coming to the problem of the CMC and its  contribution to the gluon dynamics. 

As it is  well known \cite{TVCM2} the CMC generates the non-perturbative Debye mass $M_D(T)$, connected to the CM string tension $\sigma^H(T)$, which is proportional to $T^2$

$$\sigma^H(T) = {\rm const}~ g^4 (T) T^2$$
as it was found on the lattice \cite{63*} and  non-perturbatively in the Appendix of \cite{42***}. The exact calculation of $G_3(s)$, which should give the explicit dependence on $M_D(T)$ is however difficult, and therefore one can use approximations  explained in Appendix 2.  

The inclusion of colour-magnetic interaction leads to the   generation of a  non-perturbative Debye mass $M_{D}$  for gluons and  quarks.   For gluons $M_{adj} \sim \sqrt{\sigma^H(T)}$,  
one   can  take it into account by  an  approximate expression for  3d Green function \cite{23}, which is derived in Appendix 2. 
\be
G_{3}(s)=\frac{1}{(4\pi s)^{3/2}}\sqrt{\frac{(M^{2}_{adj})s}{sinh(M^{2}_{adj})s}} \label{2.9}
 \ee 
 It should be mentioned that the  resulting gluon pressure Eq. (\ref{eqg}) is in   a good agreement with the lattice data\cite{23}.  \par

In the non-interacting case i.e. $\sigma^H=0$ and $L_{adj}=1$ one  obtains  the  ideal gas pressure:
\be
P_{gl}=P_{0}=\frac{(N_{c}^{2}-1)}{45}\pi^{2}T^{4}\label{2.10}
\ee

For quarks   one can write the expression in   the same form as  in (\ref{eqg}), but with the quark mass term $e^{-m^{2}_{q}s}$:
\be 
P_{f}=\sum_{q=u,d,s}P_{q},
P_{q}=\frac{4N_{c}}{\sqrt{4\pi}}\int^{\infty}_{0}\frac{ds}{s^{3/2}}e^{-m^{2}_{q}s}S_{3}(s)\sum_{n= 1, 2,...}(-)^{n+1}e^{-\frac{n^{2}}{4T^{2}s}}L^{n}_{f} \label{eqf}\\
S_{3}(s)=\frac{1}{(4\pi s)^{3/2}}\sqrt{\frac{(M^{2}_{f})s}{sinh(M^{2}_{f})s}}, 
M^{2}_{adj}=\frac{9}{4}M^{2}_{f},L_{f}^{n}=(L_{adj}^{n})^{4/9}\label{2.12}\ee
And again  in the case of massless non-interacting fermions one obtains:
\be
P_{f}=N_{c}N_{f}\frac{7T^4}{180}\label{2.13} 
\ee  
 The  full pressure reads as:
\be
P_{tot}=P_{f}+P_{gl}\label{2.14}
\ee

Integrating over proper time interval   ds  in (\ref{eqf})   and  replacing  the square root term in (\ref{2.12}) by an  approximate  exponential term \cite{24,26} one  obtains
 \be P_{f}=\sum_{q=u,d,s}P_{q},\frac{P_{q} (T,\mu)}{T^4} = \frac{2 N_c}{\pi^2} \sum_n \frac{(-)^{n+1}}{n^2} L^n K_2  \left( \frac{\bar M n }{T}\right) \frac{\bar M^2}{T^2}, \label{press}\ee 
 where $\bar M =  \sqrt{m^2_f + \frac{M^2(T)}{4}}, ~~ M(T) = a\sqrt{\sigma_s (T)} , a \approx 2 $   \cite{TVCM2,25}.

 To include the effects of the baryon chemical potential we should do the substitution in (\ref{eqf}):
\be
L^{n}_{f} \to L^{n}_{f}cosh(\mu n/T) \label{div}
\ee

	\section{ Polyakov line  calculations.}
	As we saw the thermodynamics of QGP in FCM is defined by two main ingredients:
the non-perturbative screening masses M(T) are calculated via $\sigma^{H}(T)$and known both 
analytically and on the lattice \cite{TVCM2, 63*}. This part is especially important at 
high T due to the growth of $\sigma^H(T)$.
Another important ingredient is the Polyakov line  L with the dynamics defined 
by the
field correlators $D^E$ and $D_{1}^{E}$ \cite{TVCM1}.     Polyakov line was introduced in \cite{3,19,20,21} as a main  dynamical ingredient of $QGP$ and it is   associated with the correlator $D_1(x)$, which produces the interaction term $V_1 (r,T)$,  with a nonzero asymptotics $V_1^E(\infty, T)$
so  that the Polyakov line was written  as $L(T) = \exp \left( - \frac{V_1(\infty, T)}{2T} \right)$. 
 
However, more  careful analysis done in \cite{64a}, has  revealed, that there are 3 sources of the Polyakov line in the non-perturbative correlators (1), two of them are due to the correlator $D^E(x)$ and one due to $ D^E_1(x)$, which also generates the perturbative part of $L(T)$. We relegate  the detailed analysis of these sources to the Appendix 3, and  ref.\cite{64a} and here we only conclude, that the main part of the  contribution of $ D_1^E(x)$ $(V_1^E (\infty, T))$  is cancelled by that of the saturated part of $D^E(x)$  and the  resulting contribution can be associated with the confining interaction of the static charge of Polyakov line with  a picked-up antiquark, which create the heavy-light system with mass $M_{HL} (T)$, so that one can continue the  previous definition of $L(T)$ as 
$$ L(T) = \exp \left( - \frac{V_1(\infty, T)}{2T}\right)\to\exp\left( -\frac{M_{HL}(T)}{T}\right).$$  

One of the ways to calculate L is
to evaluate it via the heavy-light mass $M_{HL}$ \cite{ka}.Here we are using as in \cite{ka} the mass 
$M_{HL}(T)$,which is T-dependent due to the temperature dependent string tension $\sigma^{E}(T)$,studied repeatedly on the lattice\cite{45,46,47},
with the relation $M_{HL}(T) \sim  \sqrt{\sigma^{E}(T)}$.
To find $\sigma^{E}(T)$ explicitly one can use a connection between $\sigma^{E} $
and the quark condensate $\lan \bar{q}q\ran$ found in \cite{51,**}, which can be associated  with the T-dependent  quark condensate,  since in the FCM approach the latter is produced by the scalar confinement \cite{49,50,51,**}.  Indeed, the lattice data on $\sigma(T)$ \cite{45,46,47} and $\bar{q}q(T)$ \cite{qlatt1,latt3,Baz} show  a similar behaviour.

	 We take the  CE string tension in the massless quark  limit   related to the chiral condensate \cite{51} as $|\lan \bar q q(T)\ran| =const~ (\sigma (T))^{3/2}$.
Introducing a  dimensionless parameter \(a(T)\) as $\sigma (T) = \sigma (0)  a^2 (T)$, one has  
\be |\lan \bar q q\ran (T)| = |\lan \bar q q\ran(0)| a^3 (T)\label{50}\ee.

\begin{figure}[h!]
			\centering
{\includegraphics[scale=0.25]{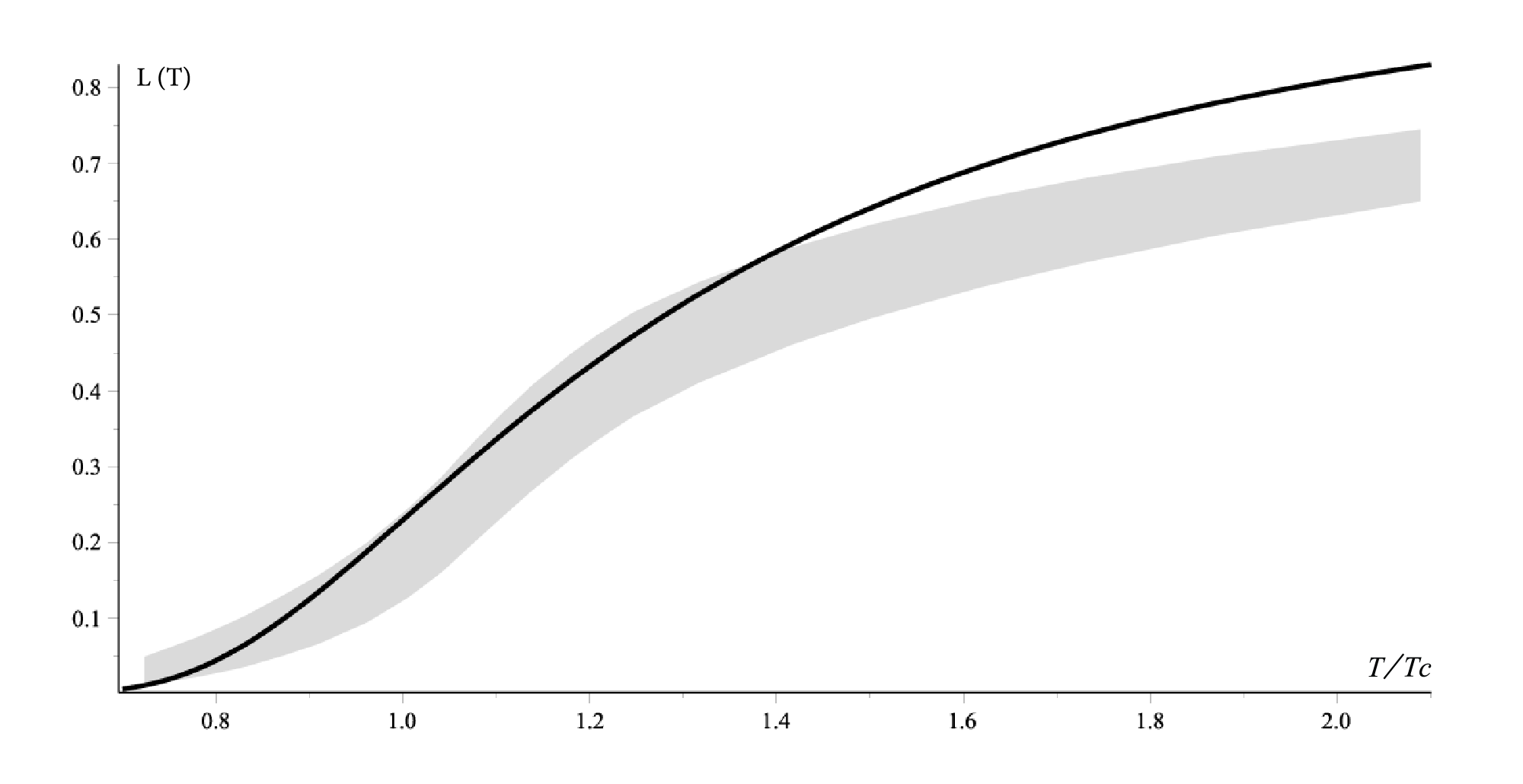}} 
									\caption{L(T) as a function of $T/T_{c}$,$T_{c}$=160 MeV.Grey band 
corresponds to $L_{HL}(T)$ within the accuracy limits of a(T). The solid black line
is the `` ideal'' $L_{FCM}(T)$ used below in the paper.} \label{FIGPol}
	\end{figure}

As a result one has $M_{HL} (T) = M_{HL} (T_0) \frac{a(T)}{a(T_0)}$ and $L(T) = \exp \left( - \frac{M_{HL} (T)}{T}\right)$.
The numerical data are shown in Fig.\ref{FIGPol}. The error band in Fig.1
corresponds to the accuracy of $a(T)$ in  the lattice data in \cite{latt3}and the solid black 
line is our  ``ideal'' FCM  line $L_{FCM}(T)$    which  on  one hand is close to  the error band, and on the other hand 
as will be seen below in the paper, yields a good agreement with lattice 
data.
    
	\begin{figure}[h!]
			\centering
{\includegraphics[scale=0.25]{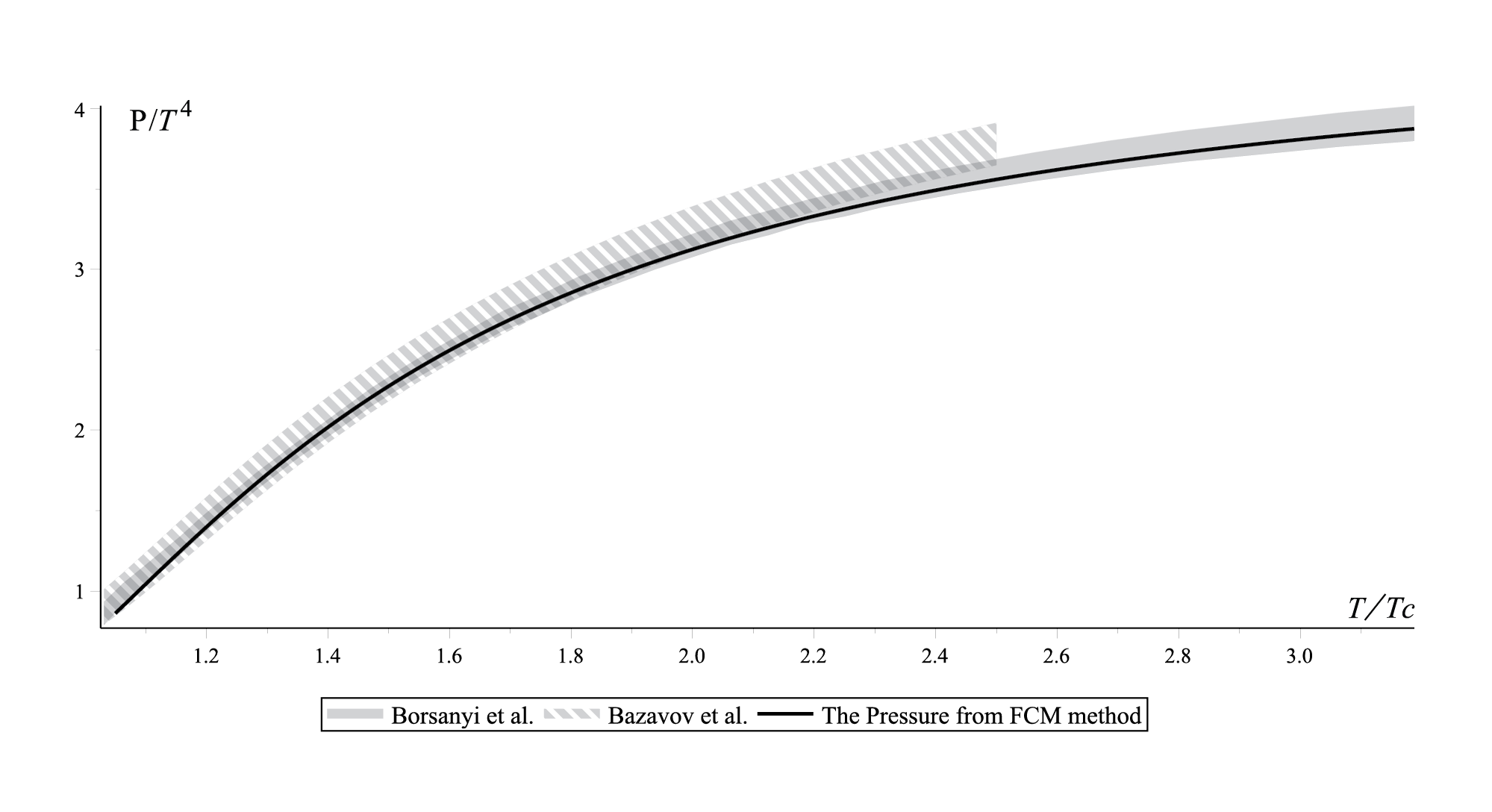}} 
									\caption{The ratio of the pressure to $T^{4}$ as a function of $T/T_{c}$. The grey band is the lattice  
 data  of Borsanyi et al.  \cite{qlatt2} and the striped band   is the lattice data from Bazavov et al. \cite{latt3}.} \label{FIGPress}
		\end{figure}
		\begin{figure}[h!]
				\centering
{\includegraphics[scale=0.25]{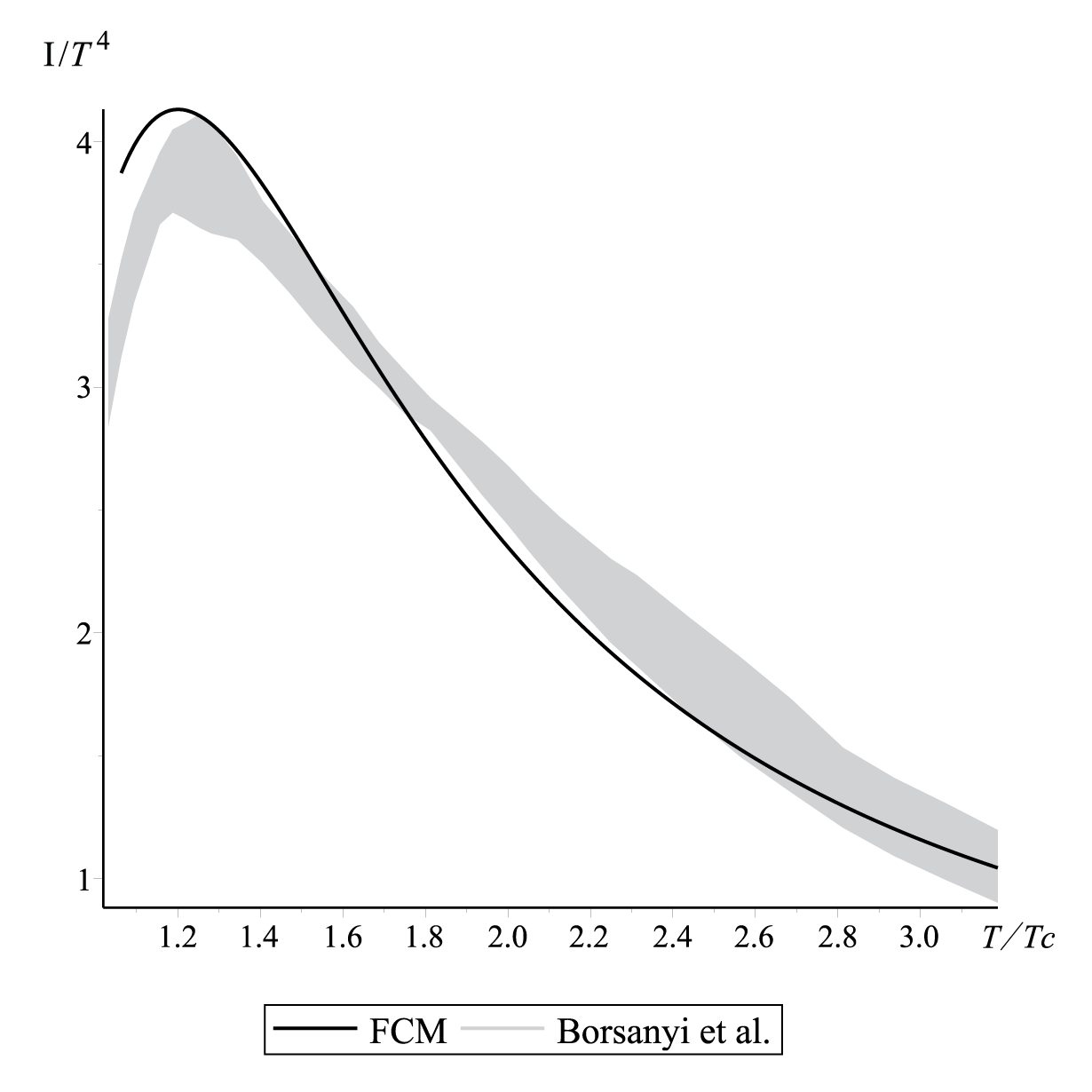}} 
									\caption{The ratio of the anomaly $(I=\epsilon-3P)$ to $T^{4}$  in QGP  as a function of $T/T_{c}$. The grey band is the lattice  data  of Borsanyi et al.  \cite{qlatt2}.} \label{Anomaly}
		\end{figure}
		
		\begin{figure}[h!]
				\centering
{\includegraphics[scale=0.25]{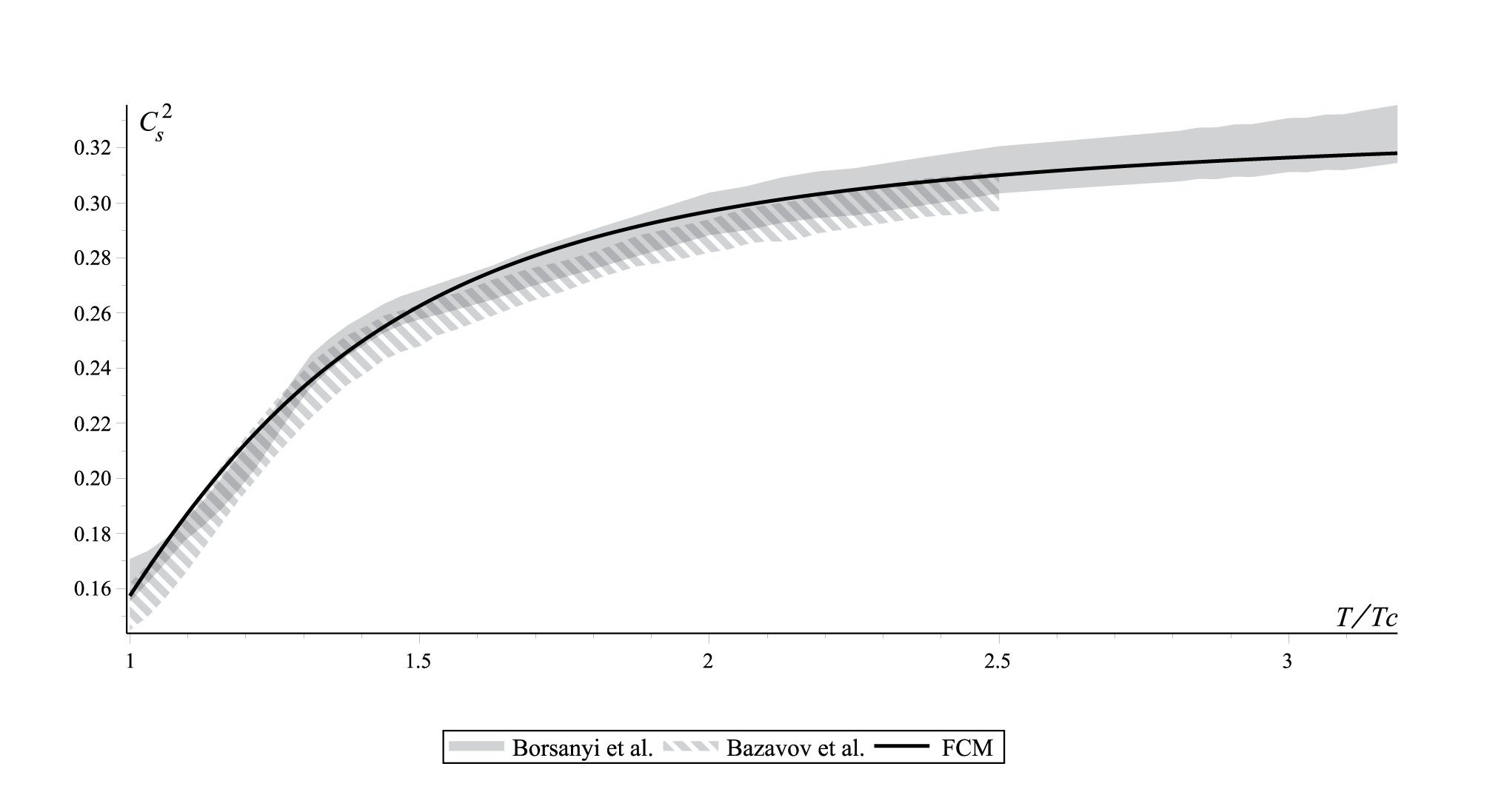}} 
									\caption{The square  of the speed of sound  in QGP as a function of $T/T_{c}$. The grey band is the lattice  data  of Borsanyi et al.  \cite{qlatt2} and the striped band   is the lattice data from Bazavov et al. \cite{latt3}.} \label{speed}
		\end{figure}

\section{The QCD thermodynamics at  finite baryon chemical potential}

One of   immediate tests of the FCM thermodynamics is the behaviour of the QGP pressure  Fig.\ref{FIGPress}, the scale anomaly Fig.\ref{Anomaly} and the speed of sound Fig.\ref{speed} , where  $L_{FCM}(T)$ and the  CMC Debye mass is taken into account, in comparison with lattice data for zero  and extension to non-zero baryon chemical potential.  One can see  that our predictions    are in good agreement  with lattice data at $\mu=0$.
 \footnote{We extended our results to rather high temperatures , just because we wanted to test our basic principles. }   \footnote{The same  is true for  the  speed of sound in gluodynamics  \cite{ourspeed}.}

At this point we extend our results  to  a finite baryon chemical potential. We will use the definition of the baryon chemical potential in the same way as in \cite{25} i.e. $\mu_{B}=3\mu_{q}$ (we will not  include  a  separate chemical potential for  the  strange quark). There is  the  possibility of comparison of our  predictions with the  Taylor expansion of $\mu$ \cite{thLatt2}. We   use the following assumption:  according to  \cite{20,71} at small densities , $\mu_B \leq 300-400$ MeV, we can neglect the influence of the baryon chemical potential on the Polyakov line.    \par

The summation over $n$ in (\ref{press})  can be done if one uses the integral representation

\be K_\nu (z) = \frac{\left(\frac{z}{2}\right)^\nu \Gamma\left(\frac12\right)}{\Gamma\left( \nu + \frac12\right)} \int^\infty_0 e^{-z\cosh t}(\sinh t)^{2\nu} dt.\label{19*}\ee

 As a result one obtains as in \cite{25}
 
\be \frac{1}{T^4} P_q (T,\mu) = \frac{N_c}{\pi^2}  (\xi_1^{(+)} + \xi_1^{(-)})\label{20*}\ee 
 with 
 \be \xi^{(\pm)}_1 = \frac13\left( \frac{\bar M}{T}\right)^4 \int^\infty_0 \frac{u^4du}{\sqrt{1+u^2}\left[ 1+\exp \left( \frac{\bar M}{T} \sqrt{1+u^2} + \frac{V_1}{2T} \pm \frac{\mu}{T}\right)\right]}.\label{21*}\ee
 Changing the integration variable, 
 \be \frac{\bar M}{T} \sqrt{1+u^2} = z+ \frac{\bar M}{T}.\label{22*}\ee
 
   The expression  (\ref{press})  can be  brought to the form: 
 \be \frac{P_q (T,\mu)}{T^4} = f_+ (T,\mu) + f_-(T,\mu),\label{3.8}\ee
 
 \be f_{\pm} (T,\mu) =\frac{N_c}{3\pi^2} \int^\infty_0 \frac{ dz\left(z^2+ 2z \frac{\bar M}{T}\right)^{3/2}}{1+\exp \left( z + \frac{\bar M}{T} +\frac{ V_{1}(T)}{ 2T} \mp \frac{\mu}{T}\right)},\label{sing}\ee
 where it is taken into account that $L=\exp \left( - V_{1}(T)/2T\right)$. 
\par
 The expression (\ref{3.8}) has no singularities at  real $\mu$, but $f^{\pm}$ may get a singularity for imaginary chemical potentials for $Im(\mu)=\pi T$
due to vanishing of the denominator in (\ref{sing}) at $z=- \frac{\bar M}{T} - \frac{V_{1}(T)}{2T}$.\par

 Hence one can conclude that in the normal situation with real  $\mu$ and $L_{f}$  the singularity in $P(\mu,T)$ is absent, this conclusion implies that there is no critical point   $T_{c}(\mu)$ in the domain of small baryon chemical potentials   and the analytic structure  is affected only by   complex singularities. From this point of view, it seems that our consideration  could be  extended without any changes to   large enough  values of the chemical potential and  temperatures   $T\le 1$   GeV   if we take $\bar M$ and  $L$ independent of $\mu$. \par

To test ourselves we  have calculated   the  pressure at $\mu_{B}=100,200,300$ MeV and  $\mu_{B}$=400 MeV. As will be  seen  in the next section there is reasonable agreement between our predictions and    lattice data,   without significant changes in QGP state with growing $\mu_B$.

\section{Results and discussion}

Below we show our results for the pressure and the sound velocity in  comparison with the lattice data. As was discussed above, we obtained
   Polyakov line expression via connection with the heavy-light meson mass  , derived from the quark condensate     using eq. (\ref{50}).  The exploited form  of the Polyakov  line  $L_{FCM}$ is shown in Fig.1 together with the  dark  region $  L _{HL}(T)$ derived from the quark condensate. One can see,  that $L_{FCM}$ is close to the $  L_{HL}$ within its accuracy region.

The data for $M_f(T)$ and $M_{adj}(T) =\frac32 M_f(T)$ are taken from the exponential approximation of the square root   expression in (\ref{2.9}),(\ref{2.12}), which was taken as   $M_f(T) = 1.6 \sqrt{\sigma^H}, $ which is near the Debye mass value, obtained in \cite{TVCM2} and ensures the  high temperature behaviour of $P(T)$, which is impossible to  reproduce without this  CMC  contribution.

One can see in Fig.2 the comparison of our FCM  result for $P(T)$ with the lattice data \cite{21,24} for the  zero baryon density. The  resulting curves coincide within  their accuracy limits.

\begin{figure}[h!]
	\centering
{\includegraphics[scale=0.2]{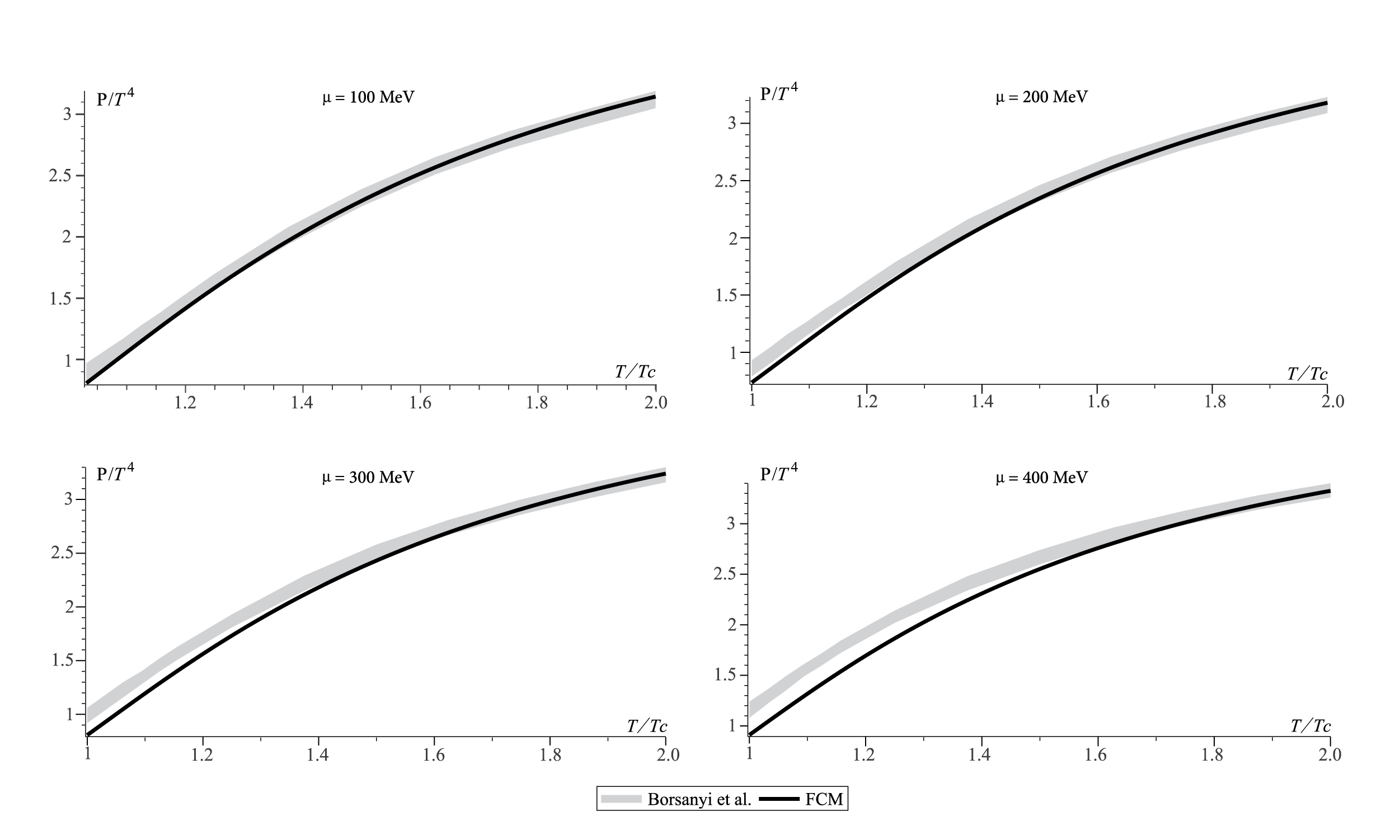}} 
									\caption{The ratio of QGP pressure to $T^{4}$ as a function of $T/T_{c}$ for  $\mu_{B}=100,200,300,400$ MeV . The grey bands are the lattice  
 data  of Borsanyi  et al. from    \cite{thLatt2} at corresponding values of $\mu_{B}$} \label{FIG1}
	\end{figure}

\begin{figure}[h!]
	\centering
{\includegraphics[scale=0.25]{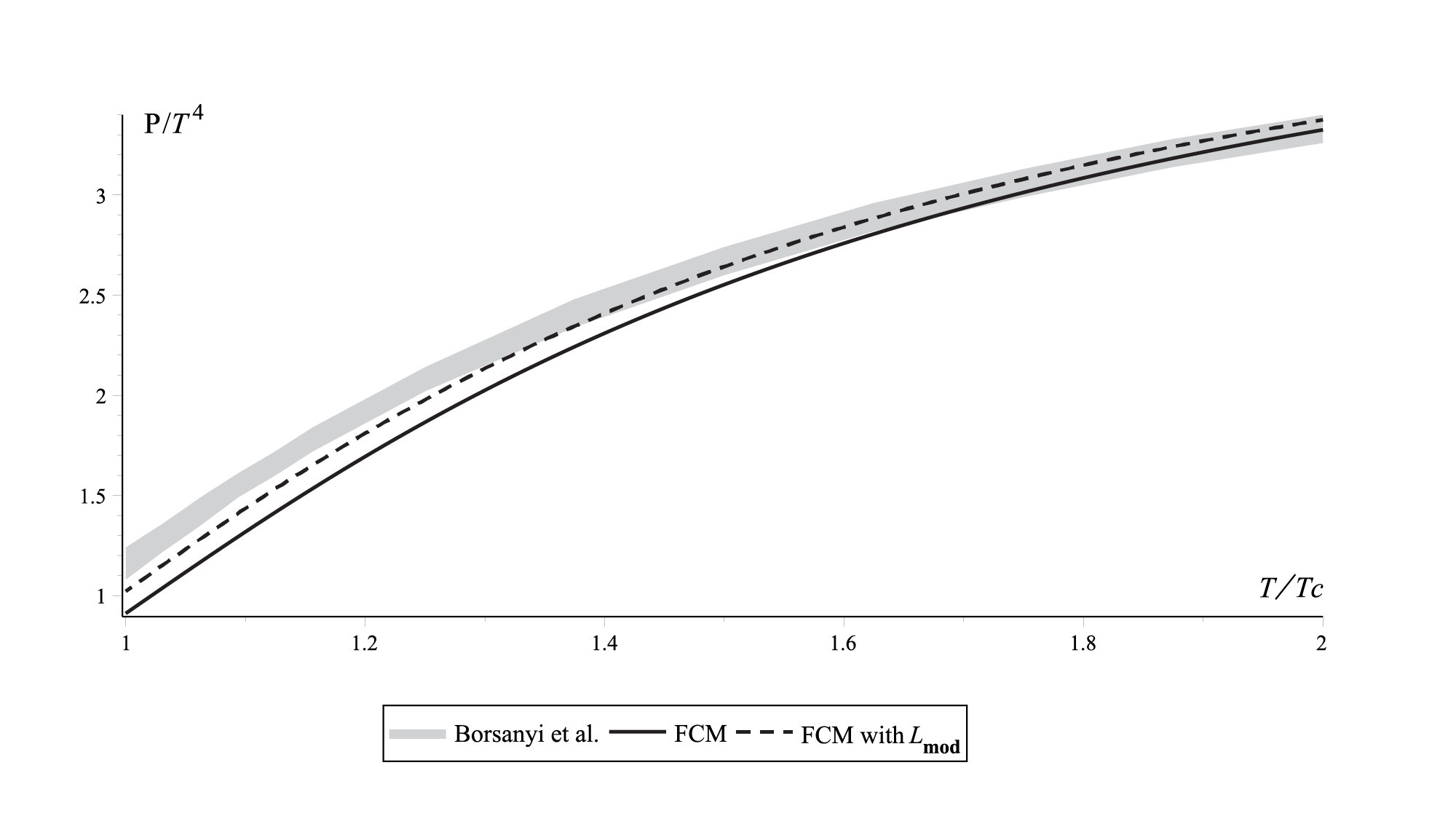}} 
									\caption{The ratio of QGP pressure to $T^{4}$ as a function of $T/T_{c}$ for  $\mu_{B}=400$ MeV  with $L_{FCM}$(black line) and  with  Polyakov line that is scaled ,similar to \cite{Polmu} (dashed line) . The grey band is the lattice  
 data  of Borsanyi  et al. from    \cite{thLatt2}.   } \label{FIG5}
	\end{figure}

Even more appealing is the agreement of our FCM results for the pressure  for  $\mu_B =0.1,0.2,0.3$ GeV in Fig.\ref{FIG1}    with the lattice  data of \cite{thLatt2} for the same values of $\mu_{B}$ . One can conclude, that at low $\mu_B$, $\mu_B\leq 0.4 $ GeV the FCM  results predict a smooth behaviour of $P(T, \mu_B)$ without any hint of a singular point and this is in agreement with  the analytic structure of $P(T,\mu_B)$ displayed in Eqs. (\ref{3.8},)  (\ref{sing}). At the same time these results agree with the similar conclusions of the lattice studies \cite{thLatt2}.The slight
disagreement with the lattice data $\mu_{B}$ = 400 MeV  (black solid line) on   Fig.\ref{FIG5}  could be connected with renormalization of Polyakov line at finite baryon densities. For example on Fig.\ref{FIG5} we  also show  the pressure(grey solid line)  with Polyakov line,  that is scaled similar to the \cite{Polmu}.\\     
One should  notice   that  the  lattice  results for the pressure in \cite{thLatt2} were  obtained in the first order of the  square of the chemical potential

\begin{figure}[h!]
	\centering
{\includegraphics[scale=0.25]{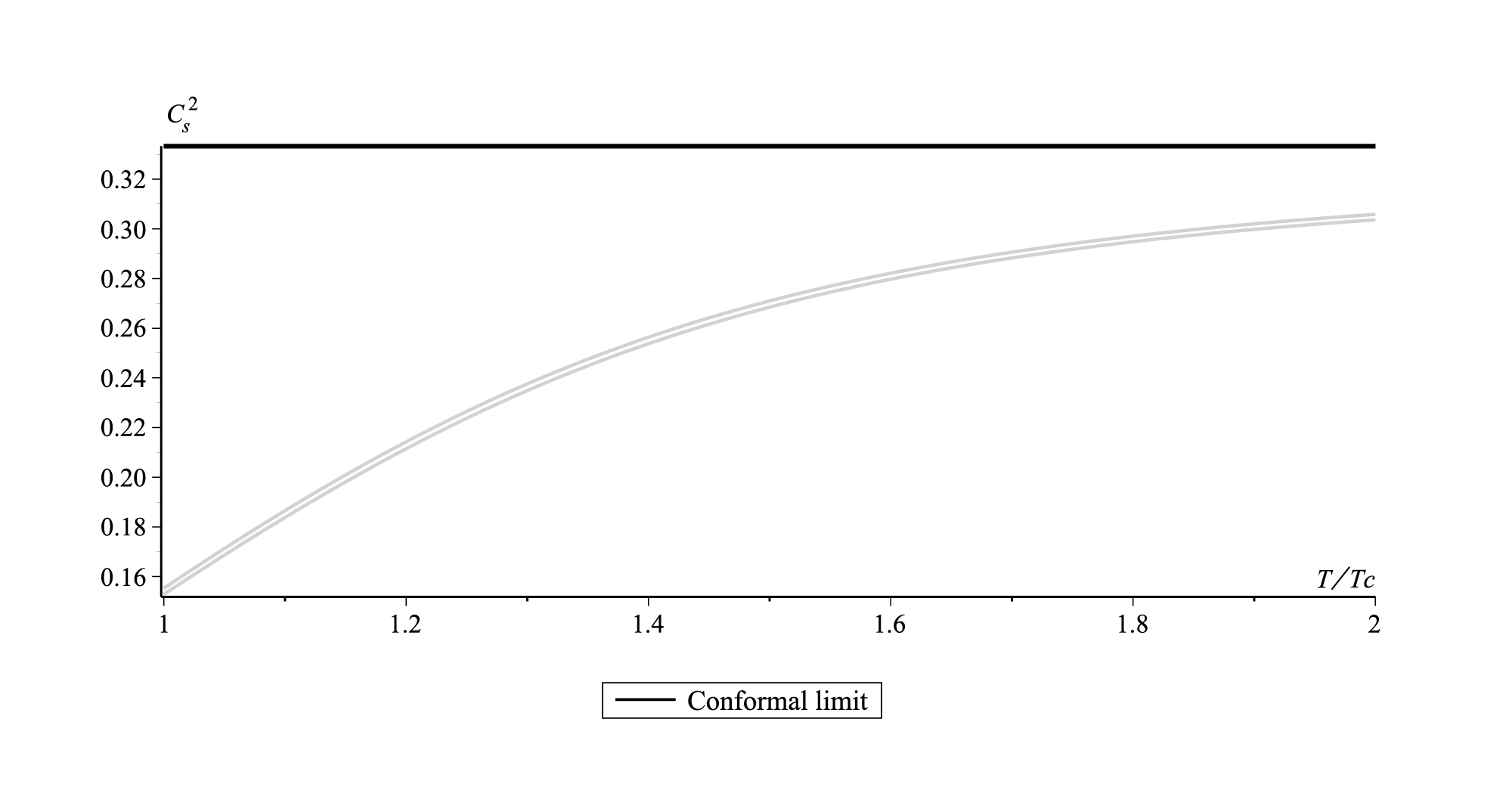}} 
							\caption{The width of  solid  line  is the changing of the square of the  speed of sound in the range  $\mu_{B}=0..300 MeV$} \label{FIG6}
		\end{figure}

  The  square of the  speed of sound, which for nonzero $\mu$ and in the isentropic  condition  can be written as (see Appendix 4 for details of derivation).    
\be
 c^2_s = \frac{ n^2\frac{\partial^2 P}{\partial T^2} - 2sn \frac{\partial^2 P}{\partial T\partial \mu}+  s^2 \frac{\partial^2 P}{\partial \mu^2}}{(\varepsilon +p) \left(\frac{\partial^2 P}{\partial T^2}\frac{\partial^2 P}{\partial \mu^2}- \left( \frac{\partial^2 P}{\partial T\partial\mu}\right)^2\right)},\label{3.2}\ee
where we have defined:
\be s=\frac{\partial  P}{\partial T },~~ n=\frac{\partial  P}{\partial \mu},~~\varepsilon +P = Ts +\mu n. \label{3.3}\ee 
 
 We show in  Fig.\ref{FIG6} the speed of sound
in the range $\mu_{B}=[0,300] $ MeV, where  the width of the line is equal to  the difference  $c^{2}_{s}(\mu_{B}=300)-c^{2}_{s}(\mu_{B}=0)$.    So from the FCM point of view  the domain  of low  chemical potentials $\mu_{B}<400 $ MeV  is safe  and could be described  by Taylor expansion  in baryon chemical potential $\mu_B$, because  in this range  this series converges, and the radius of  convergences is defined  by $\frac{\mu}{T} =  \pm i \pi$  Roberge--Weiss point in Eq. (\ref{sing}).  \par

\section{Conclusions and outlook}
    
The present paper is devoted to the   effects of small baryon chemical potential $\mu_B$ $
\leq 400$ Mev in the  dynamics of QGP.
 
 It is an extension of the study of QCD thermodynamics at vanishing baryon density    and is in the line of the series of papers \cite{3,19,20,21,22,23,24,25,26} where the QCD thermodynamics   is worked  out on the basis of FCM.\par 
  
  We have exploited above  the FCM thermodynamics to calculate the QGP pressure at finite baryon density in the   temperature range $1<T/T_c<2$, where  $T_c=160$ MeV.
  
  Our basic dynamics was  defined by two factors; the Polyakov line that is connected with  $L_{HL}(T) = \exp (-M_{HL}/T)$, and the colormagnetic confinement ( CMC) in the exponential form  with the CMC quark mass $M_D=c \sqrt{\sigma^H(T)},$ where $c =1.6$  is close to the  $q\bar q$ Debye mass in \cite{TVCM2} with $c=2$.
  
  We have used for the heavy-light mass $M_{HL}(T)$ calculated from the $T$- dependent string  tension  $\sigma^E(T)$, defined from the quark condensate  $\lan q\bar q(T)\ran$ measured on the lattice.     The Polyakov line $L_{FCM}$ exploited in the paper is close to  the accuracy limits   $L=\exp (-M_{HL} (T)/T)$  
  
  We have  demonstrated that the resulting  pressure $P_{FCM} (T, \mu)$ is in good  agreement   with lattice data of the Budapest-Wuppertal \cite{qlatt2,thLatt2} and Hot QCD groups \cite{latt3} in both cases   for zero and non-zero chemical potentials. 
	We have  also calculated  changing in the speed of the  sound that one could compare   with FiG.[7] in  \cite{thLatt2}.
  
  From this point of view our analytic equations (\ref{eqg}), (\ref{3.8}) can be   considered as  an analytic counterpart of the corresponding  lattice data. 
  
  All this implies the absence of  a critical point  in   the  studied range of  $T$ and $\mu_B$ from the point of view of FCM method.
  
  It should be noted however that we have used both $M_D$  and $M_{HL}$ independent of $\mu_B$   in the range $\mu_{B}<400$ MeV.
  
  The interesting region of high $\mu_B$, $\mu_{B}>1 $ GeV, is possibly hiding a  completely different picture, with a singular behaviour of pressure  and sound velocity, as it was found in \cite{73}. However this phenomenon is strongly connected  with a possible dependence of $L(\mu)$ and $M_D(\mu)$recently studied on the lattice in \cite{71}. These results are planned for the next paper.

\section{Acknowledgements}
We thank   M. A. 
Andreichikov and B. O. Kerbikov, M. A. Zubkov,E.A.Fedina,  R.A.Abramchuck and especially  S.I.Blinnikov for very fruitful discussions.\par 
 This work was done in the frame of the scientific project, supported by the Russian Science Foundation Grant No. 16-12-10414.

\newpage

{\bf Appendix 1.The FCM formalism in thermodynamics }  {\bf  }\\

 \setcounter{equation}{0} \def\theequation{A1.\arabic{equation}}

  To start  the FCM approach in the  thermodynamics one can consider gluons and quarks in the background vacuum fields, which can  contain both colorelectric and colormagnetic fields.  Separating perturbative and vacuum gluonic fields, $A_\mu=B_\mu+a_\mu$, one can calculate the gluon and quark propagators in the loweest order in $a_\mu$ and take their  vacuum average, as it was done in \cite{64*}. In  this way one obtains the free energy and pressure in the lowest order of the standard perturbation theory, but with the full account of the averaged vacuum fields, given by correlators $D(x)$ and $D_1(x)$ in  (1). Thus the free energy of gluons can be written via the gluon propagator in the form \cite{64*}
  $$ \frac{1}{T} F_0 (B) = \frac12 \ln \det G(B) - \ln \det (-D^2(B))=$$ 
  \be = tr \left\{ - \frac12 \int^\infty_0 \zeta (s) \frac{ds}{s} e^{-sG(B)}+ \int^\infty_0 \zeta (s) \frac{ds}{s} e^{-sD^2(B)}\right\},\label{3*}\ee where $G(B)$ is the gluon propagator and $D^2(B)$ ghost propagator
  in the background field
  \be G^{ab}_{\mu\nu} = - D^2 (B)_{ab}\delta_{\mu\nu} - 2 g F^c_{\mu\nu}(B) f^{acb},\label{4*}\ee
  \be (D_\lambda)_{ca} =\partial_\lambda\delta_{ca}- g f_{bca} B^b_\lambda,\label{5*}\ee
 while $\zeta(s)$ is the standard  regularizing factor , $\zeta(s) = \lim \frac{d}{dt} \frac{M^{2t}s^t}{\Gamma(t)}|_{t=0}$, the exact form of  it is inessential and it is not written in what follows.
 
 Our final results require the vacuum field average of (\ref{3*}) and the introduction of the temperature $T$.  The vacuum averaging is to be done with quadratic combinations of fields $B_\mu$ in the exponent in (\ref{3*}), and to this end one can use the cluster  expansion \cite{64**}.
 
 \be \lan \exp  f(B)\ran_B= \exp \left\{ \lan f (b) \ran_B + \frac12 [\lan f^2 (B)\ran_B - \lan f (B)\ran_B^2 ] +...\right\},\label{6*}\ee
which is  well  converging, as shown in \cite{VCM5} due to the  small vacuum correlation length $\lambda\la 0.2$ fm.

Next is the problem of the gluon (quark) Green's function in the external field, which can be represented as the  path integral with the phase factor, containing the external field explicitly -- this is called the Fock--Feynman--Schwinger representation (FFSR) \cite{64*}, and has  the following form in the simplest case of the ghost Green's  function
\be (-D^2)_{xy} = \lan x| \int^\infty_0 ds e^{sD^2(B)}|_y \ran = \int^\infty_0 ds (DZ)_{xy} e^{-K} \hat \Phi(x,y).\label{7*}\ee
Here
\be  K= \frac14 \int^s_0 d\tau \left( \frac{dz_\mu}{d_\tau}\right)^2, ~~ \hat \Phi(x,y) = \exp ig \int^x_y B_\mu (z) d z_\mu\label{8*}\ee 
and the integral is taken along the trajectory of the ghost $y<z_\mu (\tau) \leq x$,  and $(Dz)_{xy}$ implies the path integral
\be (Dz)_{xy} =\prod^N_{m=1} \frac{d^4\zeta(m)}{(4\pi\varepsilon)^2} \frac{d^4p}{(2\pi)^4} e^{ip (\sum_m \zeta(m) -x+y))}\label{9*}\ee
where$\zeta(m)$ is the elementary piece of the path.

In a similar way the gluon propagator has the same representation (\ref{7*}) but with additional factor in (\ref{7*}) multiplying $\hat \Phi(x,y), ~~\hat \Phi(x,y) \to \hat \Phi(x,y) \exp (-2 ig \int^s_0 d\tau \hat F_B (z(\tau)))\equiv \hat \Phi_F (x,y) $.

The next step is the introduction of the temperature $T$ within the Matsubara formalism.

In the path integral the latter implies the only replacement:$(Dz)_{xy} \to (Dz)^w_{xy}$, where the upper index $w$ means the winding path integral, which comes from $x$ to the final point $y$ for the sequence of time intervals,  $n\beta \equiv n/T$
\be (Dz)^w_{xy} = \prod^N_{M=1} \frac{d^4\Delta z(m)}{(4\pi\varepsilon)^2}\sum_{n=0,\pm 1,...} \frac{d^4p}{(2\pi)^4} e^{ip (\sum_m \Delta z(m) -(x-y)-n\beta \delta_{\mu 4})}.\label{10*}\ee
 As it is seen in (\ref{10*}) the term with $n=0$ would yield the $T$-independent contribution to the pressure, contradicting the free gluon gas result, and should be omitted in what follows. The sum over $n=\pm 1,\pm 2$ gives the twice of the sum over $n=1,2,3,..$ As it is, only closed trajectories with $x=y$ are entering in the free energy (pressure).
 
 As a result the pressure $P_{gl} V_3 =- \lan F_0 (B)\ran$ can be written as follows:
 \be P_{gl} = T \int^\infty_0 \frac{ds}{s} \frac{d^4x}{V_3} (Dz)^w_{xx} e^{-K} \left[ \frac12 tr \lan \tilde\Phi_F (x,x)\ran -\lan tr \tilde \Phi (x,x)\ran \right].\label{11*}\ee

 \newcommand{\vevar}{\mbox{\boldmath${\rm \varepsilon}$}}
 As it is clear in (\ref{11*}) the difference in the square brackets contains two effects: 1) the ghost reduction of the gluon d.o.f. 2) the gluon spin interaction corrections, since the operator $\hat F_{\mu\nu}$ entering  in $\hat \Phi_F$ has the following representation
 \be -2i \hat F_{\mu\nu} = 2 (\veS \veB^{(1)} + \veS^{(1)}\vevar^{(1)})_{\mu\nu},\label{12*}\ee
  where $\veS^{(1)}$ is the gluon spin operator and $\veB^{(1)}, \vevar^{(1)}$ are the background colormagnetic and colorelectric fields.

 Therefore, neglecting at  the first step the spin-dependent contribution one can replace the term in the square brackets simply by  the  adjoint Wilson loop, and as a result one obtains the  representation, given below in (\ref{3*}).  It is easy to understand the form  (\ref{3*}),  considering the free case with vacuum fields $B_\mu \equiv 0$. In  this case, as demonstrated in the Appendix of \cite{64*}, one obtains from the square brackets in  (\ref{11*}) $\left( \frac12 \cdot 4 -1\right)(N^2_c-1) = N^2_c-1$ and 
 \be P_{gl} (B=0)=\varphi(B=0)=(N^2_c-1) \frac{T^4\pi^2}{45}.\label{13*}\ee

 Summarizing, the final result for the gluon pressure can be written in the form Eq.(\ref{2.3}) \cite{3,21,23}.

 \vspace{2cm}

{\bf Appendix 2 }  {\bf Colormagnetic confinement contribution to $S_3(s), G_3(s)$  }\\

 \setcounter{equation}{0} \def\theequation{A2.\arabic{equation}}

As one can see in (\ref{eqg}), $G_3(s)$  ($S_3(s)$)  contains the contribution of
the adjoint   (fundamental) loops respectively, which are subject to the area
law, $\lan \hat{ tr}_i W_3\ran= \exp (-\sigma_i  {\rm are} a(W))$  $i= $ fund,
adj. Kinetic term is in $K_{3d}$ in (\ref{eqg}), so both $G_3(s)$ and $S_3(s)$
are proportional to the  Green's functions of two color charges, connected by
confining string, from one point $x$ on the loop to another (arbitrary) point,
e.g. the point  $u$ on the same loop. As it was shown in \cite{25}, one can
represent $G_3(s)$ in the  spectral sum form \be G_3(s) = \frac{1}{\sqrt{\pi
s}} \sum_{\nu =0,1,2,..} \psi^2_\nu (0) e^{-2m^2_\nu s},\label{13}\ee where
$m_\nu$ are eigenvalues of the Hamiltonian of two  adjoint  charges, connected
by   the string, and $\psi_\nu (x)$ is  its eigenfunction in 2d.

As was discussed in \cite{25}, the spectral  sum in (\ref{13}) does not converge well, especially at large $T$, therefore one should calculate the
combined effect of all terms. A simple example is given by the free case:
$\sigma_i =0$. In this  case one has

\be G_3^{(0)} (s)  = S_3^{(0)} (s)=\frac{1}{\sqrt{\pi s}} \int
\frac{d^2p}{(2\pi)^2} e^{-2p^2s} = \frac{1}{(4\pi s)^{3/2}}. \label{14}\ee

In this case one obtains the results for $P_{gl}, P_q$, which have been found
before in \cite{20}

\be P_{gl}^{(1)} = \frac{2(N_c^2-1)T^3}{  \pi^2} \sum_{n=1,2,..}\frac{
L^{(n)}_{adj}}{n^4} , ~~   P_{q}^{(1)}  = \frac{4 N_c T^4}{ \pi^2}
\sum_{n=1,2,..} (-1)^{n+1} L^{(n)}_{f}\varphi_q^{(n)},\label{15}\ee where
$\varphi_q^{(n)}$ is \be \varphi^{(n)}_q = \frac{n^2m^2_q}{2T^2} K_2 \left(
\frac{nm_q}{T}\right).\label{16}\ee

One can  see in (\ref{15}) the Stefan-Boltzmann limit -- for $L_{adj}=L_f=1$.

There are two ways, how the CM confinement can be taken into account, suggested
in \cite{25}.  Considering the oscillator interaction between the charges, one
obtains \be G_3^{OSC} (s) = \frac{1}{(4\pi)^{3/2}\sqrt{s}} \frac{M_{\rm
adj}^2}{sh M_{\rm adj}^2 s}\label{17}\ee
 and $S^{OSC}_3 (s)$ is obtained from (\ref{17}), replacing $M_{\rm adj}$, by
 $M_f$. Here $M_{\rm adj} = 2 \sqrt{\sigma_s}= m_D (T)$, where $m_D (T)$ is the
 Debye mass, calculated in \cite{TVCM2} in good agreement with lattice data.

 A more realistic form obtains, when one replaces the linear interaction
 $\sigma_s r \to \frac{\sigma_s}{2} \left( \frac{r^2}{\gamma} + \gamma\right)$,
 varying the parameter $\gamma$ in the final expressions, imitating in this way
 linear interaction by an oscillator potential. Following \cite{23}   one obtains
 \be G_3^{\rm lin} (s) = \frac{1}{(4\pi s)^{3/2}} \left( \frac{ M^2_{\rm adj}
 s}{sh (M_{\rm adj}^2 s)}\right)^{1/2}, ~~ S_3^{\rm lin} (s) = G_3^{\rm lin}
 (s) |_{M_{\rm adj}\to M_f}.\label{18}\ee

 Finally, substituting these expressions in (\ref{eqg}), (\ref{eqf}), one obtains
 the equations for $P_{gl}^{\rm lin}, P_q^{\rm lin}$, containing the effects of CM
 confinement, which  will be used in what follows. However to simplify the square-root expressions one can use for $M^2s\la 1$  the approximation with the square root term replaced by the exponential, $\left( \frac{M^2s}{shM^2s}\right)^{1/2}\approx \exp \left(-\frac{M^2s}{4}\right)$, which has a reasonable accuracy for $T<1$ GeV.

 \vspace{2cm}

{\bf Appendix 3.Nonperturbative contribution to the Polyakov line }  {\bf  }\\

 \setcounter{equation}{0} \def\theequation{A3.\arabic{equation}}

The contribution of $D^E_1(x)$ and $D^E(x)$ to the $q\bar q (gg)$ interaction can be written in terms of local potentials $V_D(r), V_D^{sat}(r), V_1^E(r), V_c(r)$ and $V_ss(r)$, which will be neglected below. Here $V_D(r) =\sigma r, ~~ V_c (r) =- \frac{4\alpha_s}{3r}$, and 

\begin{align}   
    &V_D^{\rm conf} (r,T) = 2r\int^{1/T}_0 d\nu (1-\nu T) \int^r_0   d\xi D^E(\sqrt{\xi^2 +\nu^2})\label{46}\\
    &V_D^{\rm sat} (r,T) = 2\int^{1/T}_0 d\nu (1-\nu T) \int^r_0 \xi d\xi D^E(\sqrt{\xi^2 +\nu^2})\label{45}\\
    &V_1^{\rm sat} (r,T) = \int^{1/T}_0 d\nu (1-\nu T) \int^r_0 \xi d\xi D_1^E(\sqrt{\xi^2 +\nu^2})-V_1^{\rm Coul}(r, T).\label{47}
\end{align}

The contributions of $V_1^E $ and $V_D^{sat}$ are strongly compensated as shown in \cite{64a}, so that one is left with $V_D(r)$ and $V_c(r)$, the letter is effective mostly at large $T$, when $L$ is close to  unity. Therefore one should take into account the potential $V_D(r) =\sigma r$, which gives rise to the heavy-light bound state with the mass $M_{HL}$ and $L=\exp \left( - \frac{M_{HL}}{T}\right)$. At this point it  is important to  fix  the renormalization procedure of  the contributing  confinement and  Coulomb interaction, which is similar in  the   lattice data of   \cite{69*} and in our case  and yields almost similar results for $L(T)$, as it is seen in \cite{64a}.

 \vspace{2cm}

{\bf Appendix 4 }  {\bf Derivation of eq.(\ref{3.2}) for the sound velocity at finite $\mu$ }\\

 \setcounter{equation}{0} \def\theequation{A4.\arabic{equation}}
 
 Below we are interested in the sound velocity $c_s$ at fixed isentropy $s/n_B$, $c^2_s = \left( \frac{\partial P}{\partial \varepsilon} \right)_{s/n_B}; \frac{s}{n_B} \equiv \bar s$
 \be \left.\frac{\partial P}{\partial \varepsilon }\right|_{\bar s} = \frac{\frac{ \partial P}{\partial T} dT + \frac{\partial P}{\partial \mu} d\mu}{\frac{ \partial \varepsilon}{\partial T} dT + \frac{\partial \varepsilon}{\partial \mu} d\mu}, ~~ \bar s = \frac{ \frac{\partial P}{\partial T} /\mu}{N} = const .\label{A4.1}\ee
  Taking into account the isentropic  condition
  
  $$ d \left( \frac{s}{n_B}\right) =0=  d\left( \frac{ \frac{\partial P}{\partial T}}{ \frac{\partial P}{\partial \mu} }\right)  =
 \frac{d\left(  \frac{\partial P}{\partial T}\right)  
  \frac{\partial P}{\partial \mu}
   - \frac{\partial P}{\partial T}
       d  \frac{\partial P}{\partial \mu} }
        {\left(\frac{ \partial P}{\partial \mu}\right)^2}  =$$ 
 \be
 =\frac{  \frac{\partial P}{\partial \mu} \left( \frac{\partial^2 P}{\partial T^2}    dT+  \frac{\partial^2 P}{\partial T\partial \mu }    d\mu               \right) -
   \frac{\partial P}{\partial T}\left(    \frac{\partial^2 P}{ \partial \mu \partial T} dT +  \frac{\partial^2 P}{\partial \mu^2}      d\mu\right)  
    }        {\left(\frac{ \partial P}{\partial \mu}\right)^2}  = 0\label{A4.2}\ee
   one obtains the relative change of $T$ and $\mu\div\frac{dT}{d\mu}$,

   \be \frac{dT}{d\mu} = \frac{\frac{\partial^2P}{\partial \mu^2} \frac{\partial P}{\partial T} -  
    \frac{\partial^2 P}{\partial\mu\partial T} \frac{\partial P}{\partial \mu}}{\frac{\partial^2P}{\partial T^2} \frac{\partial P}{\partial \mu}- \frac{\partial^2 P}{\partial\mu \partial T} \frac{\partial P}{T} }=\frac{a}{b}\label{A4.3}\ee

 As a result one obtains from (\ref{A4.1}) (dividing numerator and  denominator by $D\mu$)
 \be 
 c^2_s =
  \frac{
  \frac{\partial P}{\partial T} a + \frac{\partial  P}{\partial\mu} b}
  {\frac{ \partial \varepsilon}{\partial T} a 
   + \frac{\partial \varepsilon}{\partial \mu} b},
   \label{A4.4}\ee
where $a,b$ are given in (\ref{A4.3}).

Now taking into account that $s=\frac{\partial P}{\partial T}$ , $n=\frac{\partial P}{\partial \mu}, ~~ \varepsilon + P  = Ts +\mu n $, one obtains the final form, given in the text.

\be 
 c^2_s = \frac{n^2\frac{\partial^2 P}{\partial T} - 2 sn \frac{\partial^2 P}{\partial  T \partial  \mu} +s^2 \frac{\partial^2 p}{\partial \mu^2}}{( \varepsilon +P) \left( \frac{\partial^2 P}{\partial T^2}\frac{\partial^2 P}{\partial \mu^2} - \left( \frac{\partial^2 P}{\partial T \partial \mu}\right)^2 \right)}\label{A4.5}\ee


\begin{thebibliography}{99}
	\bibitem{QGP1}STAR Collaboration: J. Adams, et al.Experimental and Theoretical Challenges in the Search for the Quark Gluon Plasma: The STAR Collaboration's Critical Assessment of the Evidence from RHIC Collisions, Nucl. Phys. {\bf A 757}, 102 (2005),	arXiv:nucl-ex/0501009.

	
\bibitem{QGP2}PHENIX Collaboration, K. Adcox, et al.,Formation of dense partonic matter in relativistic nucleus-nucleus collisions at RHIC: Experimental evaluation by the PHENIX collaboration, Nucl.Phys. {\bf A 757}, 184 (2005),arXiv:nucl-ex/0410003.


\bibitem{QGP3}I. Arsene et al.,  BRAHMS collaboration,Quark Gluon Plasma an Color Glass Condensate at RHIC? The perspective from the BRAHMS experiment. Nucl.Phys.A 757:1-27,(2005),	arXiv:nucl-ex/0410020.

\bibitem{QGP4}M.Gyulassy, L.McLerran, Nucl. Phys. {\bf A 750}, 30  (2005),arXiv:nucl-th/0405013.

\bibitem{QGP5}B.B.Back et al.,  (PHOBOS), Nucl. Phys. {\bf A 757},  28 (2005),arXiv:nucl-ex/0410022.

\bibitem{HIC7}E.V Shuryak, Rev. Mod. Phys. {\bf 89}, 035001 (2017),arXiv:0807.3033v2.

\bibitem{HIC8}P. Braun-Munzinger, V. Koch, T. Schafer, and J. Stachel, Phys. Rept.{\bf  621},
76 (2016), arXiv:1510.00442.

\bibitem{HIC9}Wit Busza, Krishna Rajagopal, Wilke van der Schee,MIT-CTP/4892,	arXiv:1802.04801.

\bibitem{HIC10}R. Pasechnik, M. Šumbera, Universe {\bf {3}},  7 (2017),arXiv:1611.01533.


\bibitem{HIC1}R. D. Pisarski and F. Wilczek, Phys. Rev. {\bf D 29}, 33841  (1984).

\bibitem{HIC2} G. F. Chapline, M. H. Johnson, E. Teller, and M. S. Weiss, Phys. Rev. {\bf  D 8}, 4302  (1973).

\bibitem{HIC3} John C. Collins and M. J. Perry, Phys. Rev. Lett. {\bf 34}, 1353  (1975).

\bibitem{HIC4}T. D. Lee, Phys. Rev. {\bf D 19}, 1802, (1979).

\bibitem{HIC5} N. Cabibbo, G. Parisi, Phys. Lett. {\bf B 59}, 67 (1975).

\bibitem{HIC6}E. V. Shuryak, Sov. Phys. JETP {\bf 47}, 212 (1978), [Zh. Eksp. Teor. Fiz. {\bf 74}, 408 (1978)].


\bibitem{Rev}R.Snellings, New J.Phys.{\bf 13}, 055008 (2011),arXiv:1102.3010.

\bibitem{Hydro1}Derek A. Teaney,arXiv:0905.2433.

\bibitem{Hydro2}T.Hirano, M. Gyulassy, Phys. {\bf A 769}, 71-94 (2006),arXiv:nucl-th/0506049.




\bibitem{qlatt1}
S. Borsanyi, Z. Fodor, C.Hoelbling, S. D Katz, S. Krieg, C. Ratti,K.K. Szabo .,10.1007/JHEP {\bf  09}, 073 (2010),arXiv:1005.3508.


\bibitem{qlatt2}{S.Borsanyi, Z.Fodor, C.Hoelbling,Phys. Lett. {\bf B 730},  99-104 (2014),arXiv:1309.5258 [hep-lat].}

\bibitem{latt4}F.Karsch, J. Phys. Conf. Ser. {\bf 46}, 122-131 (2006); J.Goswanu, et al, Conference C18-07-22; arXiv:1811.02494.


\bibitem{latt1}Owe Philipsen, Prog. Part. Nucl. Phys. {\bf 70}, 55 (2013), arXiv:1207.5999.

\bibitem{latt3}A.Bazavov,T.Bhattacharya, C. DeTar et al.,Phys. Rev.{\bf D 90}, 094503 (2014),arXiv:1407.6387.

\bibitem{lattglu4}S. Borsanyi, G.Endrodi, Z.Fodor, A.Jakovac, S. D. Katz, S.Krieg, C.Ratti, K.K. Szabo, JHEP {\bf 1011}, 077,(2010),
arXiv:1007.2580.

\bibitem{25*} Y. Aoki, G. Endrodi, Z. Fodor, S.D. Katz, K.K. Szabo, Nature, {\bf 443}, 675 (2006), arXiv: hep-lat/0611014.

\bibitem{Crit}P.Parotto, M.Bluhm, D.Mroczek et al, MIT-CTP-5015,	arXiv:1805.05249.
\bibitem{Blinnikov} Tobias Fischer, Niels-Uwe F. Bastian, Meng-Ru Wu, Petr Baklanov, Elena Sorokina, Sergei Blinnikov, Stefan Typel, Thomas Klähn, David B. Blaschke,https://www.nature.com/articles/s41550-018-0583-0
\bibitem{LV}LIGO Scientific and Virgo Collaborations,
Phys.Rev.Lett. 119 (2017) no.16, 161101,
 arXiv:1710.05832 [gr-qc].
\bibitem{Itoh}N. Itoh, Prog. Theor. Phys., 44, 291-292 (1970)
\bibitem{Witten} E.Witten,Phys. Rev. D 30, 272 (1984). 

\bibitem{thLatt1}A. Bazavov, H.-T. Ding, P. Hegde et al., Phys. Rev. {\bf D 95}, 054504 (2017),arXiv:1701.04325.

\bibitem{thLatt2}Sz. Borsanyi, G. Endrodi, Z. Fodor, S. D. Katz, S. Krieg, C. Ratti, K. K. Szabo, JHEP
{\bf 1208},  053 (2012), arXiv:1204.6710.

\bibitem{thLatt3}J. Gunther, R. Bellwied, S. Borsanyi, Z. Fodor, S. D. Katz, A. Pasztor and C. Ratti, Nucl.Phys. {\bf A  967},  720 (2017),arXiv:1607.02493.
\bibitem{Kot1}V. G. Bornyakov, V. V. Braguta, E.-M. Ilgenfritz, A. Yu. Kotov, A. V. Molochkov, A. A. Nikolaev, JHEP 1803161  (2018), arXiv:1711.01869.
\bibitem{Kot2}V. V. Braguta, E.-M. Ilgenfritz, A. Yu. Kotov, A. V. Molochkov, A. A. Nikolaev, Phys. Rev. {\bf  D 94 } 114510 (2016), arXiv:1605.04090.
\bibitem{Kot3}V.V. Braguta, E.-M. Ilgenfritz, A.Yu. Kotov, B. Petersson, S.A. Skinderev, Phys. Rev. {\bf D 93},   034509 (2016), arXiv:1512.05873.
\bibitem{Kot4}V. V. Braguta, V. A. Goy, E.-M. Ilgenfritz, A. Yu. Kotov, A. V. Molochkov, M. Muller-Preussker, B. Petersson,  JHEP {\bf 1506},   094 (2015),arxiv:1503.06670.
\bibitem{VCM2}H.G. Dosch, Phys. Lett. {\bf B 190}, 177 (1987).
\bibitem{VCM3}H.G. Dosch, Yu.A. Simonov, Phys. Lett. {\bf B 205}, 339 (1988).
\bibitem{VCM4}Yu.A. Simonov, Nucl. Phys. {\bf B 307}, 512 (1988).
\bibitem{VCM5}A.Di Giacomo, H.G. Dosch, V.I. Shevchenko and Yu.A. Simonov, Phys. Rep. {\bf 372}, 319 (2002),arXiv:0007223.
\bibitem{VCM6}Yu.A. Simonov, Phys. Usp.  {\bf 39}, 313 (1996),arXiv:hep-ph/9709344.
\bibitem{VCM7} D.S. Kuzmenko, V.I. Shevchenko, Yu.A. Simonov, Phys. Usp. {\bf 174}, 3 (2004),arXiv:0310190.
\bibitem{3}Yu.A.Simonov,  Ann.  Phys.  {\bf 323}, 783 (2008), hep-ph/0702266.



\bibitem{19}E.V.Komarov,  Yu.A.~Simonov,  Ann.  Phys.   {\bf 323}, 1230 (2008),   arXiv:hep-ph/0707.0781.


\bibitem{20}  Yu.A.~Simonov, M.A.Trusov,   Phys.   Lett.  {\bf B 650}, 36 (2007),   arXiv:hep-ph/0703277.

\bibitem{21}A.V.~Nefediev, Yu.A.~Simonov, M.A.Trusov, Int. J. Mod. Phys.   {\bf E 18}, 549 (2009),   arXiv:hep-ph/0902.0125.


\bibitem{22} N.O.Agasian, M.S.Lukashov and Yu.A.Simonov, Mod. Phys. Lett. {\bf
A 31}, 1050222 (2016); arXiv: 1610.01472.
\bibitem{23} N.O.Agasian, M.S.Lukashov and Yu.A.Simonov, Eur. Phys. J. {\bf A 53}, 138 (2017); arXiv: 1701.07959.
   
    \bibitem{24}   M.S.Lukashov and Yu.A.Simonov, JETP Lett. {\bf 105}, 691 (2017);  arXiv: 1703.06666.
\bibitem{25} M.A.~Andreichikov,   M.S.~Lukashov and Yu.A.~Simonov, Int. J. Mod. Phys. {\bf A 33}, 8 (2018),
       arXiv:1707.04631.
\bibitem{26} M.A.~Andreichikov,    and Yu.A.~Simonov, Eur. Phys. J. {\bf C 78}, 5 (2018), 
       arXiv:1712.02925.

\bibitem{45}  O.Kaczmarek, F.Karsch, E.Laermann, M.Lutgemeier
,  Phys. Rev.  {\bf D 62}, 034021 (2000), hep-lat/9908010.
\bibitem{46} P.Bicudo, and N.Caroso, Phys. Rev. {\bf D 85}, 077501 (2012), arXiv: 1111.1317.

\bibitem{47}P.Cea, L.Cosmai, F.Cuteri and A.Papa, JHEP, {\bf 6}, 2 (2016);
arXiv: 1511.01783.
\bibitem{48}
A.~Bazavov, Y. Burnier  and P.~Petreczky, Nucl. Phys. {\bf A932}, 117 (2014), arXiv:1404.4267.







\bibitem{Baz}S.Borsanyi, C.Hoelbling , Z.Fodor, et al.,  POS Lattice {\bf 2010}, 185 (2014), arXiv:1011.4230.






\bibitem{49} Yu.A.Simonov, Phys. Atom. Nucl.  {\bf 60}, 2069 (1997).
\bibitem{50} Yu.A.Simonov, Phys. At. Nucl. {\bf 67}, 846 (2004); hep-ph/0302090.


\bibitem{51} Yu.A.Simonov, Phys. At. Nucl. {\bf 67}, 1027 (2004); hep-ph/0305281.
\bibitem{**} Yu.A.Simonov, Phys. Rev. {\bf D 99}, 056012 (2019); arXiv:1804.08946.

\bibitem{60*} B.S.De Witt, Phys.  ReV. {\bf 162}, 1195 (1967);\\ G.'t Hooft, Nucl. Phys. {\bf B 62}, 444 (1973).

\bibitem{60**} Yu.A.Simonov, Phys. At. Nucl. {\bf 74}, 1223 (2011), arXiv:1011,5386.
\bibitem{42***} Yu.A.Simonov, Phys. Rev.D, 096002 (2017); arXiv:1605.07060. 

\bibitem{TVCM1}Yu.A.Simonov, Phys. Lett.  {\bf B 619}, 293 (2005),arXiv:hep-ph/0502078.

\bibitem{TVCM2}N.O.Agasian, Yu.A.Simonov, Phys. Lett. {\bf B 639}, 82 (2006),arXiv:hep-ph/0604004.


\bibitem{42*} M.D'Elia, A.Di Giacomo and E.Meggiolaro, Phys. Rev. {\bf D 67}, 114504 (2003), hep-lat/0205018.


\bibitem{63*} G. Boyd, J. Engels, F. Karsch, E. Laermann, C. Legeland, M. Lutgemeier, B. Petersson ,  Nucl. Phys. {\bf B 469}, 419 (1996), G. S. Bali, J. Fingberg, U. M. Heller, F. Karsch, and K. Schilling, Phys. Rev. Lett. {\bf 71}, 3059 (1993), G.Boyd,J.EngelsF.Karsch,E.Laermann,C.Legeland,M.Lütgemeier,B.Petersson,  Phys. Lett. {\bf B 346}, 94 (1995).




\bibitem{42**}A.V.Nefediev and Yu.A.Simonov, Phys. Atom. Nucl. {\bf 71}, 171 (2008).




 

\bibitem{64a} R.A.Abramchuk, Z.V.Khaidukov and Yu.A.Simonov, arXiv:1812.01998.
\bibitem{ka}Yu.S.Kalashnikova, A.V.Nefediev , Yu.A.Simonov,	Phys.Rev. {\bf D 64} (2001) 014037,	arXiv:hep-ph/0103274

 
 
 


\bibitem{71}M.D.'Elia, F.Negro, A.Rucci and F.Sanfilippo, arXiv:1907.09461.

\bibitem{ourspeed}Z.V. Khaidukov, M.S.Lukashov, and Yu.A. Simonov,Phys. Rev. {\bf  D 98}, 074031(2018),	arXiv:1806.09407.

\bibitem{Polmu}Massimo D'Elia, Francesco Di Renzo, Maria Paola Lombardo, 	Phys. Rev. {\bf  D 76},114509,(2007) 	arXiv:0705.3814 [hep-lat]
\bibitem {73}Z.V. Khaidukov    and Yu.A. Simonov, 	arXiv:1811.08970,(2018).


 \bibitem{64*} Yu. A.Simonov, Phys. At. Nucl. {\bf 58}, 309 (1995), hep-ph/9311216.

\bibitem{64**} N.G.Van Kampen, Phys. Rep. {\bf C 24}, 171 (1976).
\bibitem{69*}A. Bazavov, N. Brambilla, H.-T. Ding,H.P Schadler,A.Vairo and J.H.Weber   Phys. Rev. {\bf D 93}, 114502 (2016).

\end{thebibliography}
\end{document}